\documentclass[preprint,3p,twocolumn]{elsarticle}

\usepackage{algorithm}
\usepackage[noend]{algpseudocode}
\usepackage{amssymb}

\usepackage{amsthm}
\usepackage{fancyvrb}
\usepackage[draft]{fixme}
\usepackage{multirow}
\usepackage{tabulary}
\usepackage{url}
\usepackage{xspace}
\usepackage{epstopdf}
\usepackage{stmaryrd}
\usepackage{rotating}
\usepackage{todonotes}

\theoremstyle{definition}
\newtheorem{mydef}{Definition}

\newcommand{\lojoin}{\includegraphics[scale=.8]{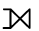}}

\begin{document}

\makeatletter
\renewcommand\paragraph{\@startsection{paragraph}{4}{\z@}%
  {-3.25ex\@plus -1ex \@minus -.2ex}%
  {1.5ex \@plus .2ex}%
  {\normalfont\normalsize\bfseries}}
\makeatother

\begin{frontmatter}

\title{On the Formulation of Performant SPARQL Queries}
%\titlerunning{On the formulation of performant SPARQL queries}

\author{Antonis Loizou\corref{cor1}}
\ead{a.loizou@vu.nl}
\author{Paul Groth}
\ead{pgroth@vu.nl}
\cortext[cor1]{Corresponding author}
%\authorrunning{A. Loizou  and P. Groth}
\address{Department of Computer Science, VU University of Amsterdam, The Netherlands.}

\begin{abstract}
The combination of the flexibility of RDF and the expressiveness of SPARQL provides a powerful mechanism to model, integrate and query data. However, these properties also mean that it is nontrivial to write performant SPARQL queries. Indeed, it is quite easy to create queries that tax even the most optimised triple stores. Currently, application developers have little concrete guidance on how to write ``good'' queries. The goal of this paper is to begin to bridge this gap. It describes 5 heuristics that can be applied to create optimised queries. The heuristics are informed by formal results in the literature on the semantics and complexity of evaluating SPARQL queries, which ensures that queries following these rules can be optimised effectively by an underlying RDF store. Moreover, we empirically verify the efficacy of the heuristics using a set of openly available datasets and corresponding SPARQL queries developed by a large pharmacology data integration project. The experimental results show improvements in performance across 6 state--of--the--art RDF stores.
\end{abstract}


\end{frontmatter}

\section{Introduction}\label{sec:intro}

Since the release of the Resource Description Framework (RDF) as a W3C Recommendation in 1999 \cite{rdf99,rdf}, the amount of data published in various RDF serialisations has been increasing exponentially. Sindice\footnote{\url{http://sindice.com}} currently indexes 15+ billion triples\cite{sindice,sindice_sparql}. The Linking Open Data cloud diagram, by Richard Cyganiak and Anja Jentzsch \footnote{\url{http://lod-cloud.net/}} provides a striking visualisation of the diversity of domains that this data covers. The query language SPARQL \cite{sparql} and SPARQL 1.1 update \cite{sparql1.1}  is the W3C Recommendation for querying RDF data . 

The flexibility in terms of both data structures and vocabularies make RDF and Linked Open Data attractive from a data provider perspective, but pose significant challenges in formulating correct, complex and performant SPARQL queries. Application developers need to be familiar with various data schemas, cardinalities, and query evaluation characteristics in order to write effective SPARQL queries. 

The contribution of this paper is a set of heuristics that can be used to formulate complex, but performant SPARQL queries to be evaluated against a number of RDF datasets. The heuristics are grounded in our experience in developing the OpenPHACTS\footnote{\url{http://www.openphacts.org}} Platform \cite{lod4pharma} -- a platform to facilitate the integration of large pharmaceutical datasets.  The efficiency of the SPARQL query templates obtained by applying these heuristics is evaluated on a number of widely used RDF stores and contrasted to that of baseline queries.

The rest of the paper is organised as follows. Section \ref{sec:motivation} gives the context and motivation for this work. A brief overview of related work is provided in Section \ref{sec:related}. Section \ref{sec:method} presents the five heuristics. Section \ref{sec:eval} provides an empirical comparison of the performance of SPARQL queries optimised using the defined heuristics. Section \ref{sec:practical} discusses the inherent difficulties in providing paginated RDF views and how these can be addressed through some of the heuristics defined in this paper.  Finally, we provide concluding remarks in Section \ref{sec:conclusions}.
 
\section{Motivation and context}\label{sec:motivation}

The work presented in this paper was carried out in the context of the OpenPHACTS project \cite{ops-ddt}, a collaboration of research institutions and major pharmaceutical companies. The key goal of the project is to support a variety of common tasks in drug discovery through a technology platform that integrates pharmacological and other biomedical research data using Semantic Web technologies. In order to achieve this goal, the platform must tackle the problem of public domain data integration in the pharmacology space and provide efficient access to the resulting integrated data. The development of the OpenPHACTS platform is driven by a set of concrete research questions presented in \cite{ops-rq}, and the platform architecture is described in \cite{lod4pharma}.

In the context of OpenPHACTS, the decision was made to avoid pushing the burden of performant query formulation to developers, but instead to provide them with a RESTful API \cite{restful} driven by parameterised SPARQL queries. This in turn created a need to formulate a set of performant queries.

A large body of work has been carried out on defining formal semantics for RDF and SPARQL in order to analyse query complexity and provide upper and lower bounds for generic SPARQL constructs
\cite{perez-acm,perez-iswc,perez-elsevier,perez-static,schmidt-foundations,schmidt-experimental,schmidt-bench,stojanovic,stocker}. These approaches are mainly focused on exploiting the formal semantics of SPARQL in order to prove generic rewrite rules for SPARQL patterns that are used in order to evaluate equivalence or subsumption between (sets of) queries. While a more detailed overview of the various SPARQL formalisation and optimisation techniques is provided in the next section, we note here that while the findings of these studies are invaluable to better understand the complexity of evaluating SPARQL queries and provide solid foundations for designing RDF store query planners and optimisers, {\em the issue of query formulation is not addressed.}

In contrast, the work presented here provides a set of heuristics to be used in formulating performant SPARQL queries based on concrete application requirements and known dataset schema. The goal is to identify patterns that can be used to formulate queries that can be effectively optimised by a wide range of RDF stores. To that end, we provide a comparison on the performance of six state--of--the--art RDF storage systems with respect to the various query formulation techniques in order to study their effectiveness and applicability. 

In summary, the paper has four main contributions:
\begin{enumerate}
\item A mapping between formal results published in the literature and SPARQL syntax.
\item A set of heuristics through which performant SPARQL queries can be formulated based on application requirements.
\item Guidance for RDF store selection based on the formulated SPARQL queries.
\item A reference set of queries and openly available datasets.
\end{enumerate}

\section{Related work}\label{sec:related}

Since the publication of the RDF: Concepts and abstract syntax \cite{rdf} W3C recommendation in 2004, a substantial body of work has been carried out by Guitierrez, P{\'e}rez, \emph{et. al.} to develop an abstract model and query language suitable to formalise and prove properties for both the RDF model and the SPARQL query language \cite{gutierrez-2004,perez-acm,perez-iswc,perez-elsevier,perez-static,schmidt-foundations}.

This section provides a selective summary of this work. The terminology given is adopted for the remainder of this paper.

In \cite{gutierrez-2004,perez-elsevier} the authors provide the following definition for RDF:
\begin{mydef}
Assume infinite sets $U$ (RDF URI references) , $B$ (Blank nodes), and $L$ (Literals). A triple $(s,p,o) \in (U \cup B) \times U \times (U \cup B \cup L)$ is an RDF triple, where $s$ is the subject, $p$ the predicate and $o$ the object. An RDF Graph $G$ is defined as a set of RDF triples. A subgraph is a subset of a graph.
\end{mydef}
The authors proceed to prove that each RDF graph contains a unique (up to isomorphism) subgraph which is an instance of $G$, denoted core(G). The $closure$ of a graph $G$ is then defined as the set of triples that can be derived (or inferred) by applying the RDFS \cite{rdfs} set of rules, denoted $cl(G)$. Thus, a \emph{normal form} for RDF graphs can be defined as $nf(G) = core(cl(G))$ and proven to satisfy two properties:
\begin{description}
\item{\textbf{Uniqueness: }} The normal form of a graph is unique.
\item{\textbf{Syntax independence: }} Let $G$ and $H$ be RDF graphs. $G \equiv H$ if and only if $nf(G) \cong nf(H)$
\end{description}
The paper concludes by proposing to eliminate redundancy in Semantic Web databases by reducing the graphs indexed to their normal form, thus reducing the number of triples that need to be considered in answering queries.

Subsequent work by the authors \cite{perez-iswc,perez-acm} provides a thorough formal study of the database aspects of SPARQL, using the definitions below.
\begin{mydef}\label{def:graph_pattern}
Assume an additional infinite set of variables $V$, disjoint from $U$, $B$ and $L$. A SPARQL graph pattern expression is defined recursively as follows.
\begin{enumerate}
\item A tuple $t \in (U \cup B \cup V) \times (U \cup V) \times (U \cup B \cup L \cup V)$ is a graph pattern. 
\item If $P_1$ and $P_2$ are graph patterns, then expressions $(P_1$ \texttt{AND} $P_2)$, $(P_1$ \texttt{\texttt{OPT}} $P_2)$, $(P_1$ \texttt{UNION} $P_2)$ are graph patterns. 
\item If $P$ is a graph pattern and $R$ is a SPARQL built--in condition, then the expression $(P$ \texttt{FILTER} $R)$ is a graph pattern. 
\end{enumerate}
\end{mydef}
\begin{mydef}\label{def:built_in}
In turn, a SPARQL built--in condition is constructed using elements of the set $U \cup B \cup L \cup V$ and constants, logical connectives($\neg$, $\wedge$, $\vee$), inequality and equality symbols ($<$, $>$, $\leq$, $\geq$, $=$), unary predicates like \texttt{isBound}, \texttt{isBlank}, \texttt{isIRI}, and more. A complete list is provided in \cite{sparql}
\begin{enumerate}
\item If $?X$, $?Y$ $\in V$ and $c\in I \cup L$, then \texttt{bound}$(?X)$, $?X=c$ and $?X=?Y$ are built--in conditions.
\item If $R_1$ and $R_2$ are built--in conditions, then $(\neg R_1)$, $(R_1\wedge R_2)$, and $(R_1\vee R_2)$ are built--in conditions
\end{enumerate}
Any graph pattern which consists of a single tuple $t$ is referred to as a triple pattern, and $var(t)$ denotes the set of variables that occur inside $t$. Similarly, for a built--in condition $R$, $var(R)$ is the set of of variables occuring in $R$.
\end{mydef}

In order to study the properties of evaluating graph patterns, the notion of a $mapping$ must also be defined.

\begin{mydef}
A mapping $\mu$ is a partial function $\mu : V \rightarrow U \cup B \cup L$. Given a triple pattern $t$, $\mu(t)$ is the triple obtained by replacing the variables in $t$ according to $\mu$. The domain of $\mu$, $dom(\mu)$ is the subset of $V$ for which $\mu$ is defined. Two mappings $\mu_1$ and $\mu_2$ are compatible when: $\forall ?X \in dom(\mu_1) \cap dom(\mu_2): \mu_1(?X) = \mu_2(?X)$. That is, $\mu_1$ and $\mu_2$ are compatible if $\mu_1$ can be extended with $\mu_2$ to obtain a new mapping.
Let $\Omega_1$ and $\Omega_2$ be sets of mappings. The following operations can be defined:
\begin{description}
\item{\textbf{Join: }}$\Omega_1 \Join \Omega_2 = \{ \mu_1 \cup \mu_2 \mid \mu_1 \in \Omega_1, \mu_2 \in \Omega_2$ and $\mu_1,\mu_2 $ are compatible$\}$
\item{\textbf{Union: }}$\Omega_1 \cup \Omega_2 = \{ \mu \mid \mu \in \Omega_1$ or $\mu \in \Omega_2\}$
\item{\textbf{Set difference: }} $\Omega_1 \setminus \Omega_2 = \{ \mu \mid \mu \in \Omega_1$,\\ \quad $\forall \mu' \in \Omega_2:$ $\mu$ and $\mu'$ are not compatible$\}$
\item{\textbf{Left outer--join: }} $\Omega_1 \lojoin \Omega_2 = (\Omega_1 \Join \Omega_2) \cup \Omega_1 \setminus \Omega_2$
\end{description}
The evaluation of a graph pattern $P$ over an RDF dataset $D$, is denoted $\llbracket P \rrbracket_D$ and is defined recursively:
\begin{enumerate}
\item If $P$ is a triple pattern $t$, then:\\
$\llbracket P \rrbracket_D = \{\mu \mid dom(\mu) = var(t)$ and $\mu(t) \in D \}$
\item If $P$ is $(P_1$ \texttt{AND} $P_2)$, then:\\
$\llbracket P \rrbracket_D = \llbracket P_1 \rrbracket_D \Join \llbracket P_2 \rrbracket_D$
\item If $P$ is $(P_1$ \texttt{OPT} $P_2)$, then:\\
$\llbracket P \rrbracket_D = \llbracket P_1 \rrbracket_D \lojoin \llbracket P_2 \rrbracket_D$
\item If $P$ is $(P_1$ \texttt{UNION} $P_2)$, then:\\
$\llbracket P \rrbracket_D = \llbracket P_1 \rrbracket_D \cup\llbracket P_2 \rrbracket_D$
\end{enumerate}
By considering the problem of deciding if $\mu \in \llbracket P \rrbracket_D$, for a given RDF dataset $D$ and graph pattern $P$, \cite{perez-acm} provides proofs for the following statements:
\begin{itemize}
\item In the general case, the evaluation of SPARQL queries is \emph{PSPACE--complete}
\item Evaluation of graph pattern expressions constructed by using only \texttt{AND} and \texttt{FILTER} operators can be solved in time $O(|P|\cdot|D|)$.
\item Evaluation of graph pattern expressions constructed by using only \texttt{AND}, \texttt{FILTER} and \texttt{UNION} operators is \emph{NP--complete}.
\item Evaluation of graph pattern expressions constructed by using only \texttt{AND}, \texttt{FILTER} and \texttt{OPT} operators is \emph{PSPACE--complete}.
\item Evaluation of graph pattern expressions constructed by using only \texttt{AND}, \texttt{UNION} and \texttt{OPT} operators is \emph{PSPACE--complete}.
\item Every graph pattern $P$ is equivalent to a pattern in \texttt{UNION} normal form:
$P \equiv (P_1$ \texttt{UNION} $P_2$ \texttt{UNION} $\cdots$ \texttt{UNION}  $P_n)$ where $\forall i: 1\leq i \leq n$, $P_i$ is constructed using only \texttt{AND}, \texttt{FILTER} and \texttt{OPT} operators.
\item Graph patterns in \texttt{UNION} normal form constructed by using only \texttt{AND}, \texttt{FILTER} and \texttt{UNION} operators can be solved in time $O(|P|\cdot|D|)$.
\item Well--designed graph patterns: The evaluation of graph pattern expressions in \texttt{UNION} normal form is coNP--complete if:
\begin{enumerate} 
\item For every subpattern of the form \\$(P$ \texttt{FILTER} $R)$, $var(R) \subset var(P)$.
\item For every subpattern of the form \\$P' = (P_1$ \texttt{OPT} $P_2)$ all variables that occur both inside $P_2$ and outside $P'$ also occur in $P_1$.
\end{enumerate}
\end{itemize}
\end{mydef}

\emph{The bounds listed above clearly indicate that the complexity of SPARQL query evaluation does not only depend on the operators used, but also on the syntactic form of the queries.} Moreover a class of graph patterns for which the evaluation problem can be solved more efficiently can be identified by imposing simple syntactic restrictions. The optimisation problem for a SPARQL query $Q$ is then framed as the process of identifying and evaluating a more efficient query that is equivalent to $Q$. Therefore in \cite{perez-static} the authors provide a set of transformation rules that can be applied to \emph{``Well--designed graph patterns''} and study the complexity of assessing containment and equivalence between SPARQL graph pattern expressions. \cite{schmidt-foundations} extends this work by proving equivalence of SPARQL queries for a total of 37 transformation rules.

A different approach to the optimisation problem is taken in \cite{stocker}. The authors propose a number of heuristics to estimate the selectivity of individual subpatterns in order to identify the most efficient order in which they should be evaluated. In this work, selectivity is defined as the fraction of triples in an RDF dataset that contain a bound subject, predicate or object in a subpattern. In the absence of summary statistics for a given dataset, it is assumed that bound subjects of a subpattern are more selective than bound objects, and bound objects more selective than bound predicates.  The authors provide empirical results using the LUBM \cite{lubm} benchmark to compare a number of models to determine the optimal ordering of subpatterns, that show significant improvements in the execution times of SPARQL queries when their constituent subpatterns are reordered based on their proposed heuristics.

\section{Heuristics}\label{sec:method}
In this section, the terminology and results of the work discussed in the previous section are reused and mapped to a set of techniques that can be applied to existing SPARQL queries and corresponding RDF datasets in order to improve their run--time performance. 

We summarise the heuristics as follows and then explain them in more detail in the following sections.
\begin{itemize}
\item {\em Minimise optional triple patterns} : Reduce the number of optional triple patterns by identifying those triple patterns for a given query that will always be bound using dataset statistics.
\item {\em Localise SPARQL subpatterns}: Use named graphs to specify the subset of triples in a dataset that  portions of a query should be evaluated against.
\item {\em Replace connected triple patterns}: Use property paths to replace connected triple patterns where the object of one triple pattern is the subject of another.
\item {\em Reduce the effects of cartesian products}: Use aggregates to reduce the size of solution sequences.
\item {\em Specifying alternative URIs}:  Consider different ways of specifying alternative URIs beyond \texttt{UNION}.
\end{itemize}

Before our detailed discussion, we define some preliminaries.  Recall that an RDF Graph $G$ is defined as a set of RDF triples $(s,p,o) \in (U \cup B) \times U \times (U \cup B \cup L)$ where $U$, $B$, and $L$ are infinite sets of URIs, blank nodes, and literals respectively. In addition, a  SPARQL graph pattern expression $P$ consists of triple patterns $t \in (U \cup B \cup V) \times (U \cup V) \times (U \cup B \cup L \cup V)$ connected with SPARQL operators and built--in conditions (Definitions \ref{def:graph_pattern} and \ref{def:built_in}). The evaluation of a graph pattern $P$ over an RDF dataset $D$, $\llbracket P \rrbracket_D$, is a set of mappings $\mu : V \rightarrow U \cup B \cup L$. Finally, $vars(t)$ is the set of variables that appear in $t$ and $\mu(t)$ is the RDF tuple obtained by replacing the variables in $t$ according to $\mu$. 

\paragraph{Assumptions}\label{sec:initial}

For ease of presentation, the heuristics described below assume a particular  style of SPARQL queries. A resource oriented approach is used, whereby each SPARQL query must return information related to a single \emph{resource}. In turn, different sets of information may be required for the same type of resource; we refer each to of these these sets as a \emph{view} of the resource. The application requirements must then specify which types of resource in the data are of interest, how many views are required for each one and a template for each view. An initial set of SPARQL queries can then be obtained by identifying graph pattern expressions that will return the types of information specified in each view template for a given resource.
%\todo{what is a view template? it needs a definition} 

Formally:
\begin{mydef}
Consider a set of RDF graphs $\mathbb{D} = \{G_1, G_2, \dots , G_m\}$ , a set of resource types $\mathbb{R} = \{r_1, r_2, \dots , r_n\}$ with istances in $U$, and a set of view templates $views(r_i) = \{v_1, v_2, \dots , v_o\}$ associated with each type. View templates are defined operationally: for each pair $(r_i,v_j) \in \mathbb{R} \times views(r_i)$ there exists a SPARQL graph pattern expression $P(r_i,v_j)$ such that the mappings in $\llbracket P(r_i,v_j) \rrbracket_\mathbb{D}$ can be used to instantiate the template $v_j$ with information that corresponds to an instance of $r_i$. $P$ is the conjuction of sub--patterns, $P_k \sqsubseteq P$,  which encode the shortest path between an instance of $r_i$ and each element of $v_j$ in the schema of graphs in $\mathbb{D}$.
\end{mydef}

The heuristics are in principle also applicable to other forms of queries, however the above assumption allows us to guarantee the termination of the algorithm proposed and to provide succinct definitions. We now look at each heuristic in detail.

\subsection{Mimimize optional triple patterns}\label{sec:optional}

Since real world datasets will often contain missing values, view templates must also allow for optional elements. As shown in the literature, SPARQL graph pattern expressions given by the conjunction of triple patterns and built--in conditions can be evaluated in $O(|P|\cdot|D|)$ time, where $|P|$ is number of triple patterns in the query and $|D|$ is the number of RDF tuples in a dataset $D$. However, by adding the OPT operator, evaluation complexity becomes coNP--complete for well--designed graph pattern expressions. It is thus desirable to minimise the number of optional elements in a view template (and the corresponding SPARQL graph pattern expression) while ensuring that every instance of $r_i$ that appears in $D$ can be used to instantiate the templates in $views(r_i)$.

To proceed, we need to introduce some additional terminology.  
\begin{mydef}
There exists a non--empty subset of core information types for each view template $v_j$, denoted $core(v_j)$. An instance of a resource $r_i$ is defined with respect to $v_j$ iff $\llbracket P(r_i,v_j) \rrbracket_\mathbb{D}$ contains a mapping for all elements of $core(v_j)$. Thus triple patterns in $P(r_i,core(v_j))$ do not occur inside an OPT operator. The remaining triple patterns $t_o \in P(r_i,v_j \setminus core(v_j))$ are optional iff the set of mappings $\llbracket P(r_i,core(v_j)) \rrbracket_\mathbb{D}$ is strictly larger than $\llbracket P(r_i,core(v_j))$ AND $t_o \rrbracket$
\end{mydef}
Algorithm \ref{alg:opt} is used to identify which sub--patterns in $P(r_i,v_j)$ should appear inside OPT operators.
\begin{algorithm}
\small
\begin{algorithmic}[1]
\caption{Identify optional sub--patterns in $P(r_i,v_j)$}
\label{alg:opt}
\Require $P(r,v)$: a graph pattern expression using only the \texttt{AND} operator.
\State $required = P(r,core(v))$
\State $queue = P(r,v) \setminus P(r,core(v))$;
\State $optional = []$
\While{$\neg$isEmpty($queue$)}
\ForAll{triple patterns $t_i \in queue$}
\If{\big(vars($t_i$) $\cap$ vars($required$) $\big)\not\equiv \emptyset$}
\State remove($queue, t_i$)
\If{$\llbracket$`\texttt{ASK}\{ $required$ \texttt{MINUS} \{$t_i$\} \}'$\rrbracket_\mathbb{D}$}
\State $optional[] = t_i$
\Else
\State $required[] = t _i$
\EndIf
\Else
\For{$j=1 \to j=\mid optional \mid$}
\If{\big(vars($t_i$) $\cap$ vars($optional[j]$) $\big)\not\equiv \emptyset$}
\State remove($queue, t_i$)
\If{$\llbracket$`\texttt{ASK}\{ $optional[j]$ \texttt{MINUS} \{ $t_i$\} \}'$\rrbracket_\mathbb{D}$}
\State $optional[j] = optional[j]$ \texttt{OPTIONAL} \{ $t_i$ \}
\Else 
\State $optional[j] = optional[j]$ ` . ' $t_i$ 
\EndIf
\EndIf
\EndFor
\EndIf
\EndFor
\EndWhile
\State $sparql = $ `\texttt{SELECT * WHERE \{}'
\ForAll{triple patterns $t_r \in required$}
\State $sparql^+ = `t_r$  .'
\EndFor
\ForAll{graph patterns $P_o \in optional$}
\State $sparql^+ = `$\texttt{OPTIONAL} \{ $P_o$ \}'
\EndFor
\State $sparql^+ = $ `\}'
\State\Return $sparql$
\end{algorithmic}
\end{algorithm}

The algorithm uses a graph pattern constructed using only AND operators, $P(r_i,v_j)$ and its core subpattern $P(r_i,core(v_j))$ as input, and iterates through all triple patterns in $P(r_i,v_j) \setminus P(r_i,core(v_j))$. We refer to triple pattern that do not appear inside OPT operators as $required$, and say that a two triple patterns are $connected$ if they share one or more variables. If $t_i$, the triple pattern under consideration, is connected to a required pattern a boolean ASK query is constructed to assess whether all triples in the dataset that match the required triple patterns also match $t_i$. If so, $t_i$ can be considered required. If not it is used to create an $optional$ graph pattern. Alternatively if $t_i$ is only connected to an already created optional pattern $P_o$, the later is replaced with $P_o$ AND $t_i$ if all triples that match $P_o$ also match $t_i$, or with $P_o$ OPT $t_i$ if they don't. Unconnected triple patterns are ignored until the next iteration and the algorithm terminates when all triple patterns in $P(r_i,v_j)$ have been characterised as required or optional. The algorithm will always terminate if the initial graph pattern encodes paths in the dataset schema that originate from the same node as described in the previous section. 

The use of this algorithm ensures that SPARQL queries will contain the minimal number of \texttt{OPTIONAL} triple patterns.

\subsection{Use named graphs to localise SPARQL subpatterns}\label{sec:named}

The run--time performance of any SPARQL query has a positive correlation to the number of RDF triples it is evaluated against. Named graphs provide an effective way to specify a subset of triples in a dataset that should be considered in evaluating subpatterns in a SPARQL query. 
\begin{mydef}
Assume that each graph $G_i$ in a RDF dataset $\mathbb{D} = \{G_1, G_2, \dots , G_m\}$ is assigned a unique identifier, name($G_i$)$\in U$. Then any SPARQL graph pattern expression $P$ can be expressed as the conjunction of subpatterns $P_i \sqsubseteq P$ such that $\forall i: \llbracket P_i \rrbracket_\mathbb{D} \equiv \llbracket P_i \rrbracket_{G_i}$.
\end{mydef}
This approach is most intuitive when each RDF graph in $\mathbb{D}$ originates from a different source, and SPARQL queries are used to collate together information from the different sources. As each source typically uses a different schema from the others, identifying which graph pattern expressions should appear inside each named graph becomes trivial. At the same time an arbitrary RDF dataset may be split into an infinite number of graphs, as two distinct graphs may contain an identical RDF triple. Thus, we do not provide an explicit algorithm to split a  RDF dataset into named graphs, but postulate that query performance is inversely proportional to the number of common variables across separate named graphs. 

Embedding graph patterns inside named \texttt{GRAPH} clauses can allow RDF store optimisers to consider a smaller set of triples in evaluating individual subpatterns. Thus, the localisation of SPARQL subpatterns in this manner is expected to reduce the complexity of the evaluation problem, and result in performance improvements.

Section \ref{sec:ex:optimise} provides a comparative study on the performance of queries obtained through the application of different combinations of the two heuristics discussed so far.

\subsection{Replace connected triple patterns with sequence paths}\label{sec:chains}

Property paths are a feature introduced in SPARQL 1.1 that specify a route through a graph between two nodes. This feature has mainly received negative attention from the community, e.g. \cite{property_paths}, as the ability to specify paths of arbitrary length makes the evaluation problem intractable in many cases. Here, we consider how a particular type of property path, \emph{Sequence Paths}, can instead be used to improve the performance of a SPARQL query.
\begin{mydef}
A sequence path expression of length 1 is a triple pattern. The conjunction of two triple patterns, $t_i = (s_i, p_i, o_i)$  and $t_j=(s_j, p_j, o_j)$ such that $o_i \equiv s_j$, can be rewritten as a sequence path $p_{i,j}$ of length 2 using the `/' operator: \\$p_{i,j} = (s_i, p_i/p_j, o_j)$. Moreover, two sequence paths can be merged together if the object of one is equivalent to the subject of the other.
\end{mydef}
Thus, one variable is eliminated with each triple pattern embedded in a sequence path. In turn, by reducing the dimensionality of the mapping sets obtained when evaluating graph (sub--)patterns a reduction in the cost of subsequent operations such as joining mapping sets or identifying unique mappings is achieved. Section \ref{sec:ex:sequence} provides empirical evidence to support the claim that replacing connected triple patterns with sequence paths can provide performance improvements.

\subsection{Reduce the effects of cartesian products}\label{sec:cartesian}
%\todo{I don't understand what this heuristic does. What's the two-line summary?}
\begin{mydef}
The evaluation of a graph pattern $\llbracket P \rrbracket_\mathbb{D}$ is used to generate a \emph{solution sequence} which is provided as the result of executing the corresponding SPARQL query. Each individual solution in the sequence is a set that contains at most one mapping per variable which appears in the query, and individual mappings may appear in more than one solution. Therefore, the number of solutions in a sequence is given by the product of the number of mappings obtained for each variable in the SPARQL query.
\end{mydef}

For example, let: 
$\llbracket P \rrbracket_\mathbb{D}=\{?s \rightarrow \_$\texttt{:foo}$ ,\, ?p \rightarrow $\texttt{rdfs:label}$ ,\, ?o \rightarrow $``\texttt{foo}''$ ,\, ?o \rightarrow $``\texttt{bar}''$ \}$. \\
Then, the corresponding solution sequence will consist of two elements: \\
$(?s \rightarrow \_$\texttt{:foo}$ ,\, ?p \rightarrow $\texttt{rdfs:label}$ ,\, ?o \rightarrow $``\texttt{foo}''$)$ and \\
$(?s \rightarrow \_$\texttt{:foo}$ ,\, ?p \rightarrow $\texttt{rdfs:label}$ ,\, ?o \rightarrow $``\texttt{bar}''$)$ 

That is, the single mappings for the $?s$ and $?p$ variables are repeated to form a solution for each of the two mappings for $?o$. This property is often perceived by end users as duplicating information in the results and in turn introduces an expensive post-processing step for applications that consume and present the results to users. SPARQL 1.1 \cite{sparql1.1} introduces a set of 7 aggregates that combine groups of mappings for the same variable: \texttt{SUM}, \texttt{MIN}, \texttt{MAX}, \texttt{AVG}, \texttt{COUNT}, \texttt{SAMPLE}, and \texttt{GROUP\_CONCAT}.  \texttt{GROUP\_CONCAT} is of particular interest in this context, as it can be applied to a group of mappings for the same variable to return a single mapping which contains the string concatenation of all values in the group. With respect to the example above, one can apply the  \texttt{GROUP\_CONCAT} aggregate to variable $?o$ to obtain the singleton solution sequence:
$(?s \rightarrow \_$\texttt{:foo}$ ,\, ?p \rightarrow $\texttt{rdfs:label}$ ,\, ?o \rightarrow $``\texttt{foo, bar}''$)$

Aggregates can thus be used to eliminate perceived duplication in result sequences, and obtain a succinct result format.

\subsection{Specifying alternative URIs}\label{sec:multiple}
The SPARQL specification \cite{sparql} recommends the use of the \texttt{UNION} keyword as a means of matching one or more alternative graph patterns. This is achieved by reusing the same variable names in two (or more) graph patterns, and mappings for these variables are derived from any of the matching graph patterns, regardless of their compatibility. For example, consider the following query:\\
\footnotesize
\texttt{PREFIX ex: <http://www.example.org\#> \\
PREFIX rdfs: <http://www.w3.org/2000/01/rdf-schema\#> \\
SELECT ?label WHERE \{ \\
\hspace*{0.5cm}\{ex:123 rdfs:label ?label\} \\ 
\hspace*{0.5cm}UNION \\
\hspace*{0.5cm}\{ex:456 rdfs:label ?label\} \\
\}}
\normalsize

The evaluation of this query will then contain mappings for the variable \texttt{?label} matching either of the two triple patterns; i.e. the mappings will include labels associated with either \texttt{ex:123} or \texttt{ex:456}. However, the fact that two alternative resource URIs have been used can not be inferred from the query results alone; one must also have access to the original query. Even so, it is not possible to discern which of the results apply to each of the resources, in effect discarding the provenance of the results.

At the same time SPARQL provides an additional two keywords that can be used to specify alternative URIs and circumvent the above issue: \texttt{FILTER} and \texttt{VALUES}\footnote{The \texttt{VALUES} keyword has been introduced to replace the \texttt{BINDINGS} keyword at a very late stage in the W3C recommendation process. The RDF stores considered in this paper however all implement \texttt{BINDINGS} and not \texttt{VALUES}. We assume that no significant difference in performance exists between the two keywords and expect RDF store vendors to match the SPARQL 1.1 specification in the near future. To remain consistent with the published specification the \texttt{VALUES} keyword is used in the paper, even though the queries run were written using \texttt{BINDINGS}.}. The example query above can thus be re-written as follows:\\

Using \vspace*{0.2cm}\texttt{FILTER}:\\
\footnotesize
\texttt{PREFIX ex: <http://www.example.org\#> \\
PREFIX rdfs: <http://www.w3.org/2000/01/rdf-schema\#> \\
SELECT ?label WHERE \{ \\
\hspace*{0.5cm} ?s dfs:label ?label \\ 
\hspace*{0.5cm} FILTER (?s = ex:123 || ?s = ex:456)\\
\}}
\normalsize
\vskip 0.2cm
Using \vspace*{0.2cm}\texttt{VALUES}:\\
\footnotesize
\texttt{PREFIX ex: <http://www.example.org\#> \\
PREFIX rdfs: <http://www.w3.org/2000/01/rdf-schema\#> \\
SELECT ?label WHERE \{ \\
\hspace*{0.5cm} VALUES ?s \{ex:123 ex:456\}
\hspace*{0.5cm} ?s dfs:label ?label \\ 
\}}
\normalsize

Both cases will generate mappings for the variable \texttt{?s} which can in turn be used to identify which resource each mapping for \texttt{?label} refers to. In addition to this, Section \ref{sec:ex:alternatives} provides empirical evidence that both \texttt{FILTER} and \texttt{VALUES} outperform \texttt{UNION} for most of the RDF stores considered.

It is therefore important to consider the performance of each of the three options to specify alternative URIs, in the context of the system being developed.
\section{Evaluation}\label{sec:eval}

An empirical evaluation was carried out to measure improvements obtained through the application of the heuristics presented in the previous section across a number of state--of--the--art RDF stores. Section \ref{sec:setup} provides details on the experimental setup used, while Section \ref{sec:experiments} describers the various experiments performed and presents their results. The entire experimental setup, including all datasets, queries and associated scripts is available online\footnote{\url{http://few.vu.nl/~alu900/perf_sparql.tar.gz}}.

\subsection{Experimental setup}\label{sec:setup}

\subsubsection{Hardware}\label{sec:hardware}

To ensure that the experiments could be completed in reasonable time and that RDF stores were able to deliver their best performance, the experiments were run using fairly powerful hardware:
\begin{itemize}
\item CPU: 2 $\times$ Intel 6 Core Xeon E5645 2.4GHz
\item RAM: 96GB RAM 1333Mhz
\item Hard drive: 4.3TB RAID 6 (7 $\times$ 1TB 7200rpm)
\end{itemize}

\subsubsection{Datasets}\label{sec:datasets}

Some of the main datasets considered by the OpenPHACTS platform were used to carry out the evaluation of this work, since the work of gathering application requirements and mapping them to SPARQL graph patterns had already been carried out. They are:

\begin{itemize}
\item ChEMBL v13\footnote{\url{https://www.ebi.ac.uk/chembl/}} RDF conversion\footnote{\url{https://github.com/egonw/chembl.rdf}}.
\item ChemSpider\footnote{\url{http://www.chemspider.com/}} and ACD Labs\footnote{\url{http://www.acdlabs.com}} Predicted Properties RDF conversion.
\item Drugbank RDF conversion provided by the Bio2Rdf\footnote{\url{http://bio2rdf.org/}} project.
\item Conceptwiki\footnote{\url{http://ops.conceptwiki.org/}} RDF conversion provided on request.
\end{itemize}

The above datasets mainly describe two types of resource: chemical compounds and targets (e.g. proteins). Additionally, a third type of resource is the interaction between a compound and a target. In total, the data contains 168 783 592 triples, 290 predicates and are loaded in 4 separate named graphs (one per dataset).

\subsubsection{RDF stores}\label{sec:stores}

Table \ref{tab:stores} lists the RDF stores used in the evaluation, along with the maximum memory usage measured during the experiments.
\addtocounter{footnote}{1}
\footnotetext{\url{http://virtuoso.openlinksw.com/}}
\addtocounter{footnote}{1}
\footnotetext{\url{http://www.systap.com/bigdata.htm}}
\addtocounter{footnote}{1}
\footnotetext{\url{http://www.ontotext.com/owlim}}
\addtocounter{footnote}{1}
\footnotetext{\url{http://www.openrdf.org/}}
\addtocounter{footnote}{1}
\footnotetext{\url{http://4store.org/}}
\addtocounter{footnote}{-4}

\begin{table}{}
\centering
\begin{tabular}{|l|c|c|}
\hline
			&		& Maximum \\
RDF Store	&Version& memory  \\
			&		& use   \\
\hline
\hline
Virtuoso 	& 07.00.3202 	& 4.4GB \\ 
Enterprise\footnotemark[\value{footnote}]					&		&	\\
\hline
Virtuoso	& 6.1.6.3127 	& 9.6GB \\ 
Open Source\footnotemark[\value{footnote}]\addtocounter{footnote}{1}			&		&	\\
\hline
bigdata\footnotemark[\value{footnote}]\addtocounter{footnote}{1}
			& 1.2.2			& 1.8GB \\
\hline
OWLIM-Lite\footnotemark[\value{footnote}]\addtocounter{footnote}{1}
			& 5.2		& 1.8GB \\ 
\hline
Sesame Native & 2.6.10	& 2.2GB \\
Java Store\footnotemark[\value{footnote}]
			& 			&	\\
\hline
Sesame		& 2.6.10	& 50GB  \\
In-memory Store\footnotemark[\value{footnote}]\addtocounter{footnote}{1}
			& 			&	\\
\hline
4store\footnotemark[\value{footnote}]\addtocounter{footnote}{1}
			& 1.1.5		& 17GB	\\
\hline
\end{tabular}
\caption{RDF stores used, corresponding version number and maximum memory use.\label{tab:stores}}
\end{table}

The only changes made to the default configurations of the RDF stores was to set a maximum memory limit to 90GB and disable any inferencing. Each store was restarted prior to running an experiment, and the following query issued as a warm--up:\\
\footnotesize
\texttt{SELECT ( COUNT ( DISTINCT * ) AS ?count )\\
WHERE \{ \\
\hspace*{0.5cm}?s ?p ?o \\ 
\}}
\normalsize

However, in the case of OWLIM-Lite the warm--up query consistently caused the internal database to become corrupted. The same behaviour was observed in trying to run the experiments without a warm--up. In fact only queries for a very small number of triple patterns executed correctly after the data had finished loading.

A random sample of 500 compounds and 500 targets was used to instantiate the SPARQL queries used in each experiment. We reiterate that targets and compounds map to the main concepts described by these datasets.

\subsection{Experiments}\label{sec:experiments}

This section provides details on how each experiment was carried out and presents the results obtained. Note that the results figures are better viewed in colour.

\subsubsection{Minimising \texttt{OPTIONAL} patterns and localising queries via named graphs}\label{sec:ex:optimise}

The first set of experiments considered four SPARQL query templates that correspond to the most frequently used OpenPHACTS API methods:
\begin{itemize}
\item Compound Information: Retrieve information about a compound.
\item Compound Pharmacology: Retrieve information about a compounds interactions with targets.
\item Target Information: Retrieve information about a target.
\item Target Pharmacology: Retrieve information about a targets interactions with compounds.
\end{itemize}

First, a baseline `Initial Query' was derived for each method using only \texttt{AND} operators (and the known graph patterns), and the performance of this query measured in milliseconds for each of the 500 applicable resources in the sample. In the figures provided in this section, the green $\times$ (first datapoint from the left) is the mean response time for the baseline. The error bars indicate the maximum and minimum values.

The baseline query was then rewritten to organise graph patterns inside named graphs, as described in Section \ref{sec:named}, to obtain a corresponding `Graph' query. The mean response time for this query is given by the magenta $\times$ (second datapoint from the left) in the figures below. As above, minimum and maximum response times for `Graph' queries are given by the corresponding error bar.

A third query, `Naive Optional', is also derived from the baseline query. Recall the definition for the core of a view template from Section \ref{sec:optional}. In a `Naive Optional' query, any graph patterns that do not retrieve instances of the core information types defined in a view template appear inside an \texttt{OPTIONAL} clause. This is depicted as the third datapoint from the left (green $\times$) in the figures provided in this section.

The fourth datapoint from the left (cyan $\times$) corresponds to the mean response time obtained by executing an `Optimised Optional' query for each of the 500 resources in the sample. Such queries are obtained through the application of Algorithm \ref{alg:opt} (Section \ref{sec:optional}) to an `Initial Query'

The final data point (green $\times$) in each of the figures in this section represents a `Graph Optional' query. These queries are obtained by organising graph patterns that appear in the corresponding `Optimised Optional' query inside \texttt{GRAPH} blocks.

\begin{sidewaysfigure*}
\centering
\includegraphics[width=\textwidth]{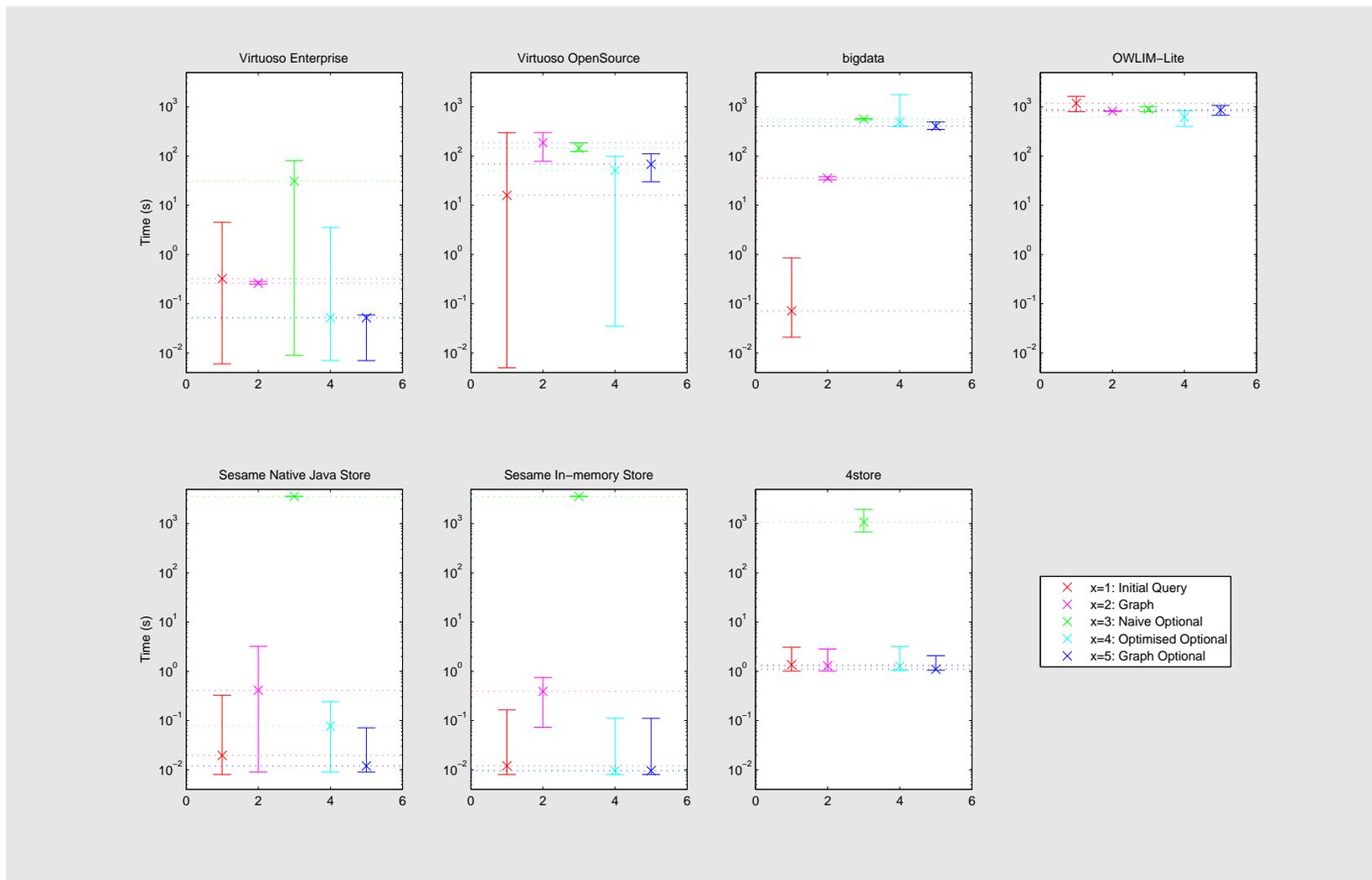}
\caption{Compound information. The average response time for the sample of 500 compounds is marked with $\times$. The error bars indicate the maximum and minimum response times obtained. \label{fig:compound_info}}
\end{sidewaysfigure*}

\paragraph{Compound Information}
Figure \ref{fig:compound_info} provides the results obtained by evaluating queries that correspond to the compound information API method across the seven RDF stores. OWLIM-Lite consistently became corrupted after 15 minutes. 

The performance improvement observed by minimising optional triple patterns is dramatic for all RDF stores with the exception of bigdata. Both of the Sesame RDF stores failed to evaluate the `Naive Optional' queries within the 30 minute timeout, while `Optimised Optional' has a response time in the order of 0.1 seconds for the Native Java Store and 0.01 for the In--memory Store. Similar improvements are obtained on average for both Virtuoso Enterprise and 4store, while the effect is still present but less pronounced for Virtuoso OpenSource.

Comparing the `Initial Query' response times against those obtained with the `Graph' queries, one can observe that introduction of named graphs actually has a negative effect on query performance for all RDF stores except Virtuoso Enterprise for which a small improvement is obtained on average. 

However, comparing the performance of `Optimised Optional' queries against that of `Graph Optional' queries reveals that the introduction of named graph for queries which do contain \texttt{OPTIONAL} patterns yields a large improvement on average for the Sesame Native Store. Considering the maximum response time measured, significant improvements are also obtained for Virtuoso Enterprise and bigdata. 

\begin{sidewaysfigure*}
\centering
\includegraphics[width=\textwidth]{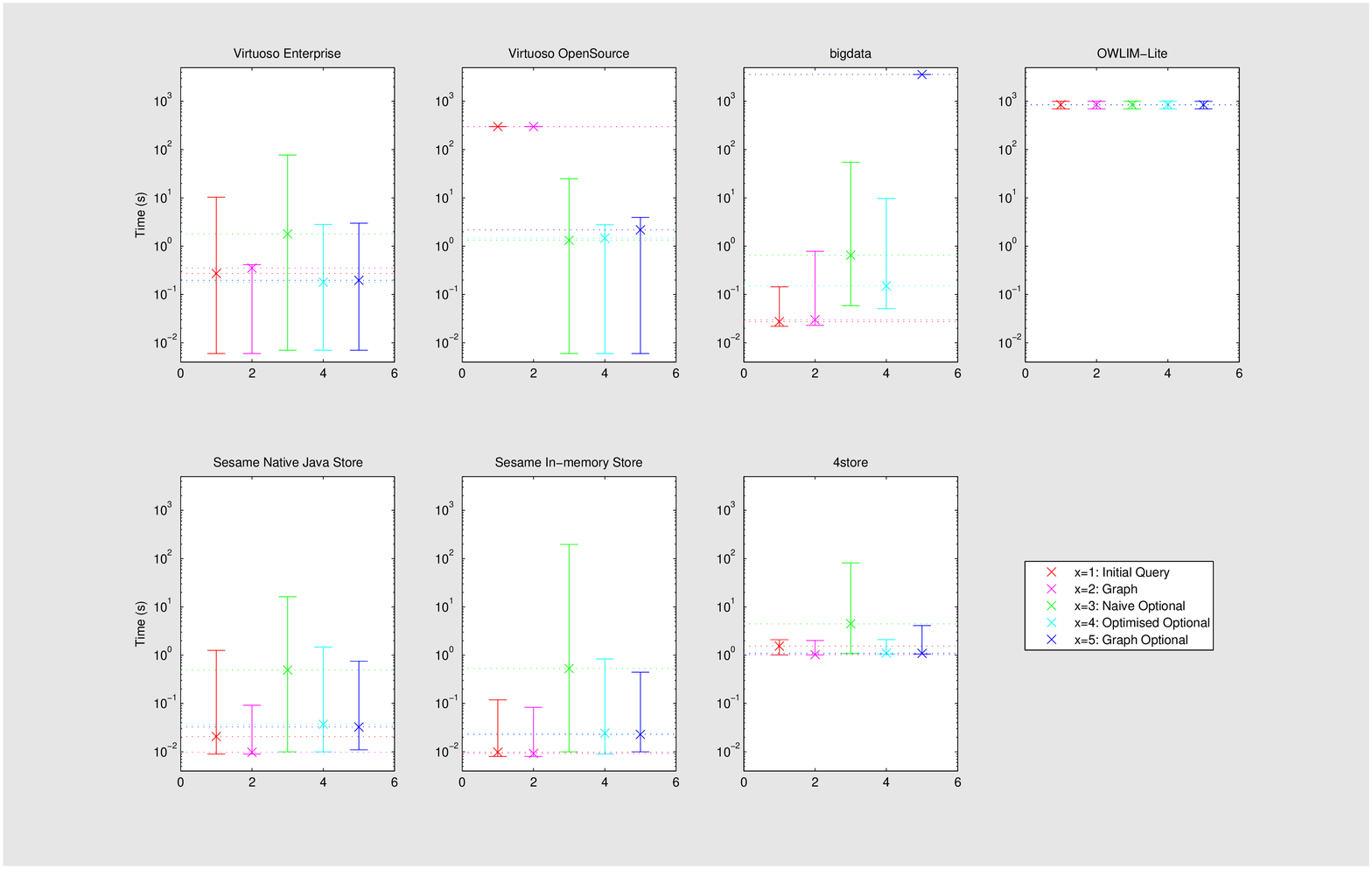}
\caption{Compound pharmacology. The average response time for the sample of 500 compounds is marked with $\times$. The error bars indicate the maximum and minimum response times obtained. \label{fig:compound_pharma}}
\end{sidewaysfigure*}

\paragraph{Compound Pharmacology}
Response times measured for queries that correspond to the compound pharmacology API method are presented in Figure \ref{fig:compound_pharma}. 

In this case, the introduction of named graphs is more effective as the maximum response time for the `Graph' query is an order of magnitude smaller for Virtuoso Enterprise and Sesame Native Java Store.  More subtle improvements are also observed for the Sesame In--memory store and 4store, while the performance of the two queries is identical for Virtuoso and an increase on the maximum response time is observer for big data. As before, no meaningful results could be obtained for OWLIM-Lite since it consistently produced an error after approximately 15 minutes.

Similarly to the previous experiment, using Algorithm \ref{alg:opt} to identify which patterns to place inside \texttt{OPTIONAL} clauses results in an improvement in average response time of over an order of magnitude for Virtuoso Enterprise and both Sesame stores, and smaller (but significant) improvements for 4store and bigdata. In contrast, for Virtuoso OpenSource the optimised queries perform slightly worse than the naive ones on average, however the upper bound on the response time of `Optimised Optional' queries is an order of magnitude less.

No significant improvements in response times were observed by introducing named graphs to the 'Optimised Optional' query. In fact, bigdata failed to evaluate the `Graph Optional' query within the 30 minute timeout interval. 

\begin{sidewaysfigure*}
\centering
\includegraphics[width=\textwidth]{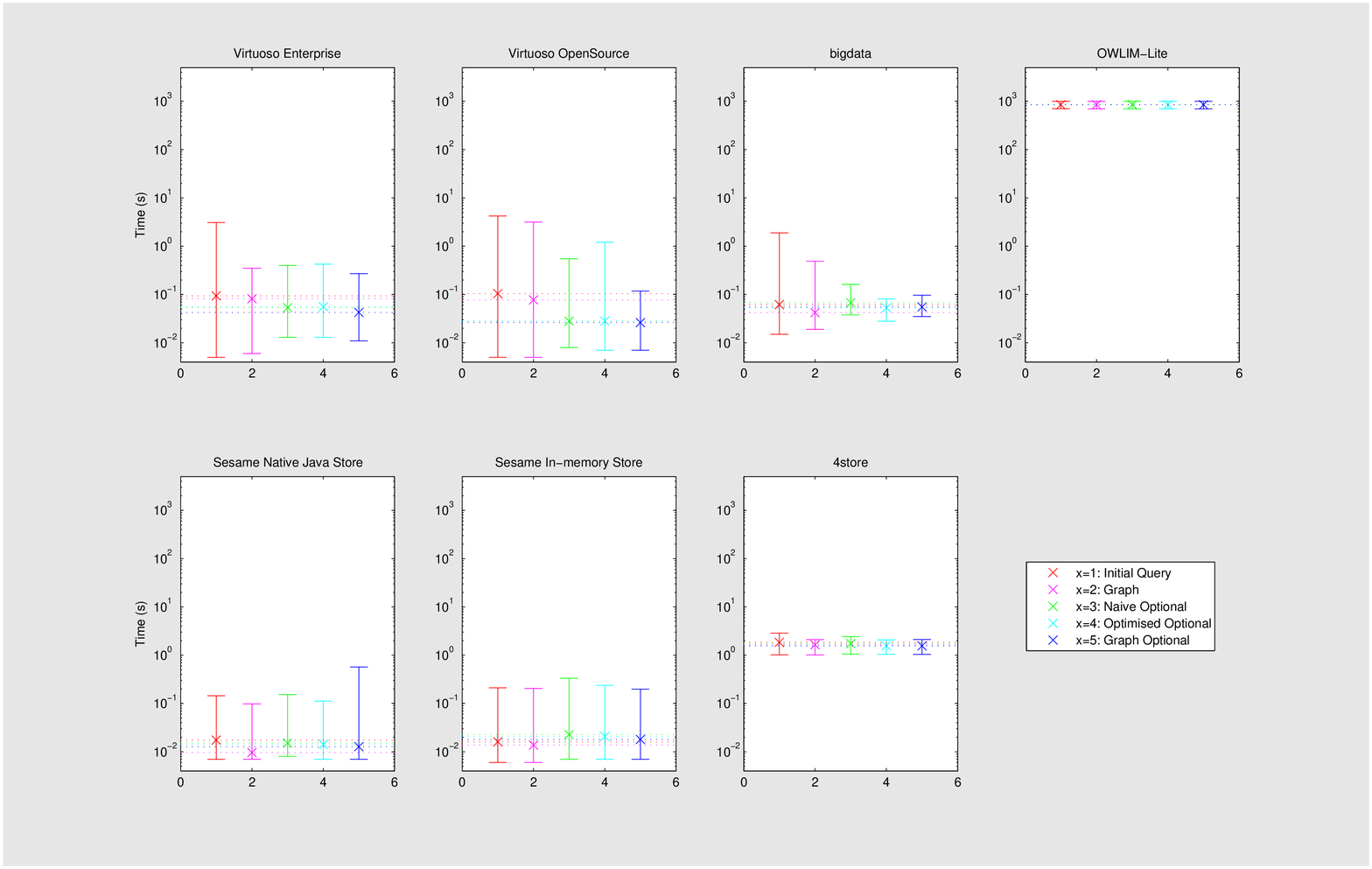}
\caption{Target information. The average response time for the sample of 500 targets is marked with $\times$. The error bars indicate the maximum and minimum response times obtained. \label{fig:target_info}}
\end{sidewaysfigure*}

\paragraph{Target Information}
Figure \ref{fig:target_info}  presents results obtained by evaluating queries that correspond to the `Target Information' method provided by the OpenPHACTS API. Overall the results of this experiment give relatively stable response times on average for all queries with respect to each RDF store (with the exception of OWLIM--Lite). However, a significant reduction of the upper bound in response time is obtained for Virtuoso OpenSource by introducing named graphs to the `Optimised Optional' query.

\begin{sidewaysfigure*}
\centering
\includegraphics[width=\textwidth]{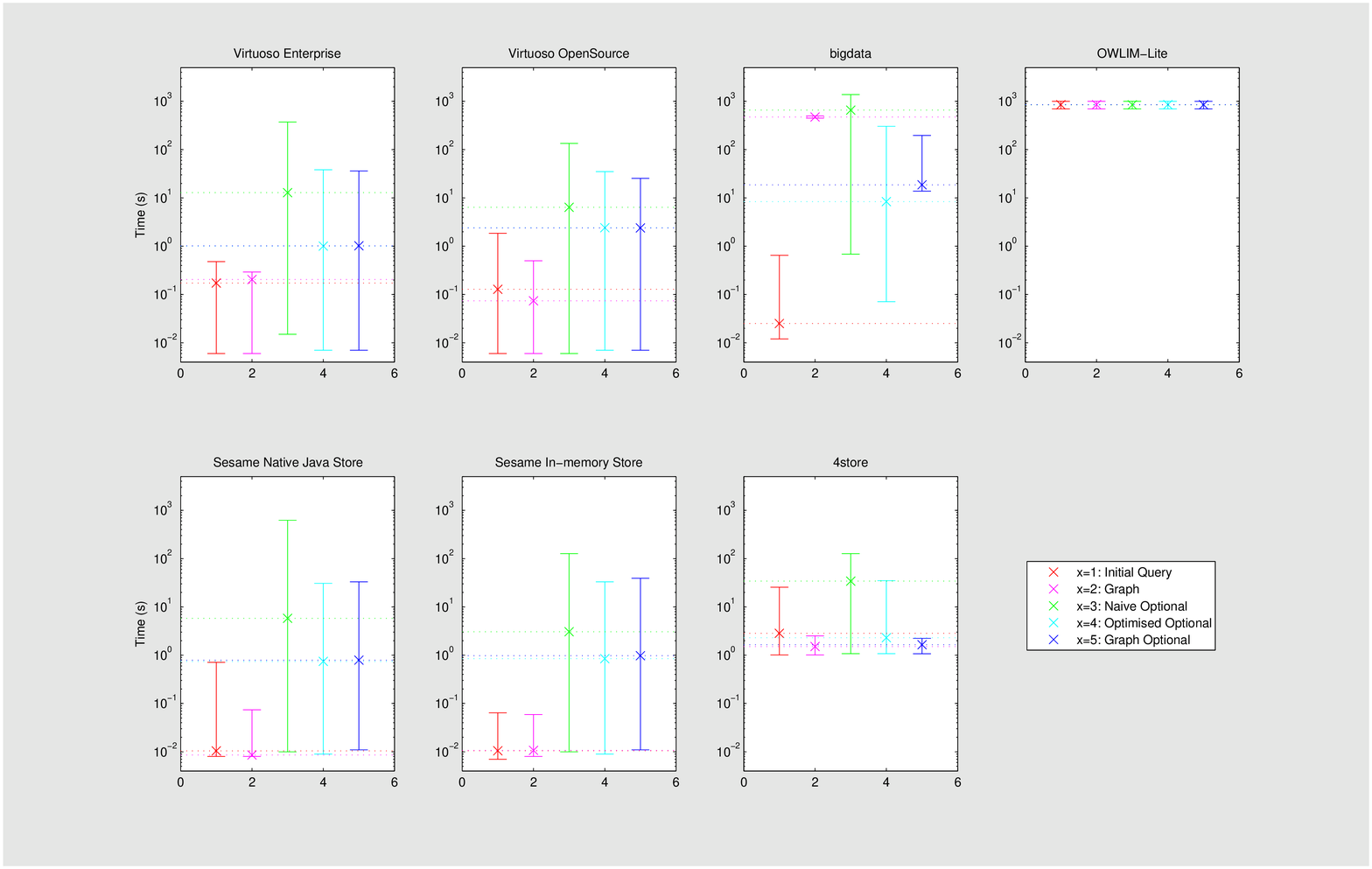}
\caption{Target pharmacology. The average response time for the sample of 500 targets is marked with $\times$. The error bars indicate the maximum and minimum response times obtained. \label{fig:target_pharma}}
\end{sidewaysfigure*}

\paragraph{Target Pharmacology}
Figure \ref{fig:target_pharma} presents the results for the fourth query considered in this experiment `Target Pharmacology'. 

In this case, the introduction of named graphs to the `Initial Query' does not yield on a performance improvement on average, but gives a significant reduction in the maximum response time for 4store and the Sesame Native Java Store. Similarly, comparing the performance of the `Optimised Optional' query to that of `Graph Optional'  we observer only an improvement with respect to the maximum response time, and only for 4store.

In contrast, the `Optimised Optional' query significantly outperforms `Naive Optional' across all RDF stores that were able to evaluate the queries. The improvement is particularly striking for bigdata, as `Naive Optional' has an average response time of just under 10 minutes, while `Optimised Optional' can be evaluated within 10 seconds on average.

\subsubsection{Sequence paths}\label{sec:ex:sequence}

\begin{figure*}
\centering
\includegraphics[width=0.6\textwidth]{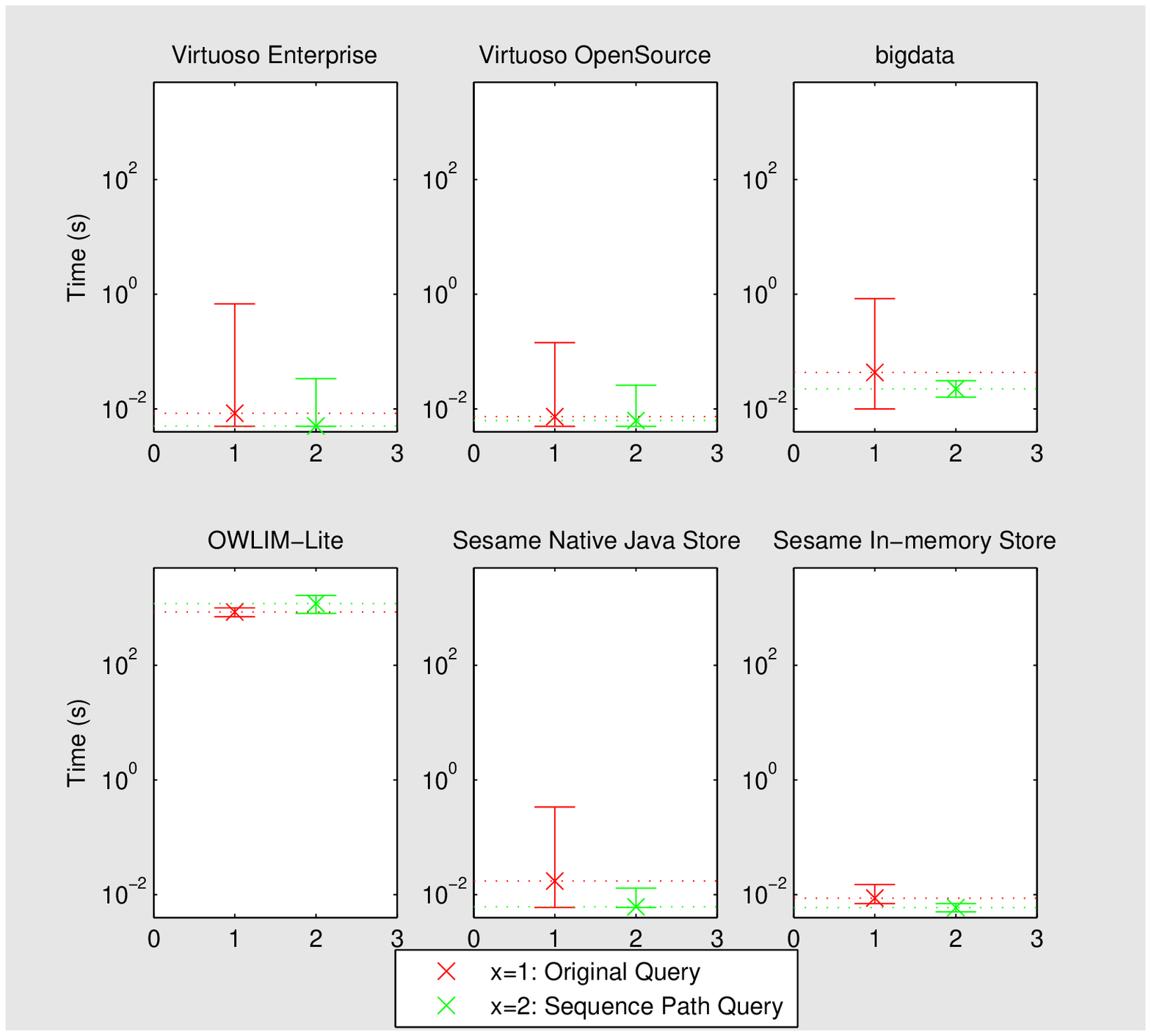}
\caption{Sequence paths. The average response time for the sample of 500 targets is marked with $\times$. The error bars indicate the maximum and minimum response times obtained. \label{fig:sequence}}
\end{figure*}

A second experiment was carried out to establish whether replacing connected triple patterns with sequence paths results in reduced response times. To do so, we compare the performance of the following two queries

Original query:\\
\footnotesize
\texttt{
PREFIX db: <http://www4.wiwiss.fu-berlin.de/drugbank\\
\hspace*{1.5cm} /resource/drugbank/>\\
PREFIX rdfs: <http://www.w3.org/2000/01/rdf-schema\#>\\
PREFIX skos: <http://www.w3.org/2004/02/skos/core\#>\\
SELECT ?synonym ?cellularLocation \{\\
\hspace*{0.5cm} [RESOURCE] skos:exactMatch ?chembl\_uri ;\\
\hspace*{1cm} rdfs:label ?synonym .\\
\hspace*{0.5cm} OPTIONAL \{ [RESOURCE] skos:exactMatch ?db\_uri .\\
\hspace*{0.5cm} ?db\_uri db:cellularLocation ?cellularLocation .\}\\
\}\\
}
\normalsize

Sequence path query:\\
\footnotesize
\texttt{PREFIX db: <http://www4.wiwiss.fu-berlin.de/drugbank\\
\hspace*{1.5cm} /resource/drugbank/>\\
PREFIX rdfs: <http://www.w3.org/2000/01/rdf-schema\#>\\
PREFIX skos: <http://www.w3.org/2004/02/skos/core\#>\\
SELECT ?synonym ?cellularLocation \{\\
 {[}RESOURCE{]} skos:exactMatch/rdfs:label ?synonym .\\
OPTIONAL \{ \\
{[}RESOURCE{]} skos:exactMatch/db:cellularLocation \\
\hspace*{2.5cm} ?cellularLocation .\}\\
\}\\
}
\normalsize

The schemas of the datasets did not allow the creation of a meaningful sequence path query for compounds. Thus, only the sample of 500 targets was used for this experiment.

Figure \ref{fig:sequence} presents the results obtained from this experiment. At the time of performing the experiments sequence paths had not been implemented in 4store, while OWLIM--Lite consistently became corrupted as in the previous experiment.

While only slight improvements are observed with respect to the mean response time, the maximum response time is significantly reduced for all four remaining RDF stores. In fact, the upper bound for the `Sequence Path' query is lower than the average response time for the `Original Query' for the two Sesame stores and bigdata.

\subsubsection{Specifying alternative URIs}\label{sec:ex:alternatives}

\begin{sidewaysfigure*}
\centering
\includegraphics[width=\textwidth]{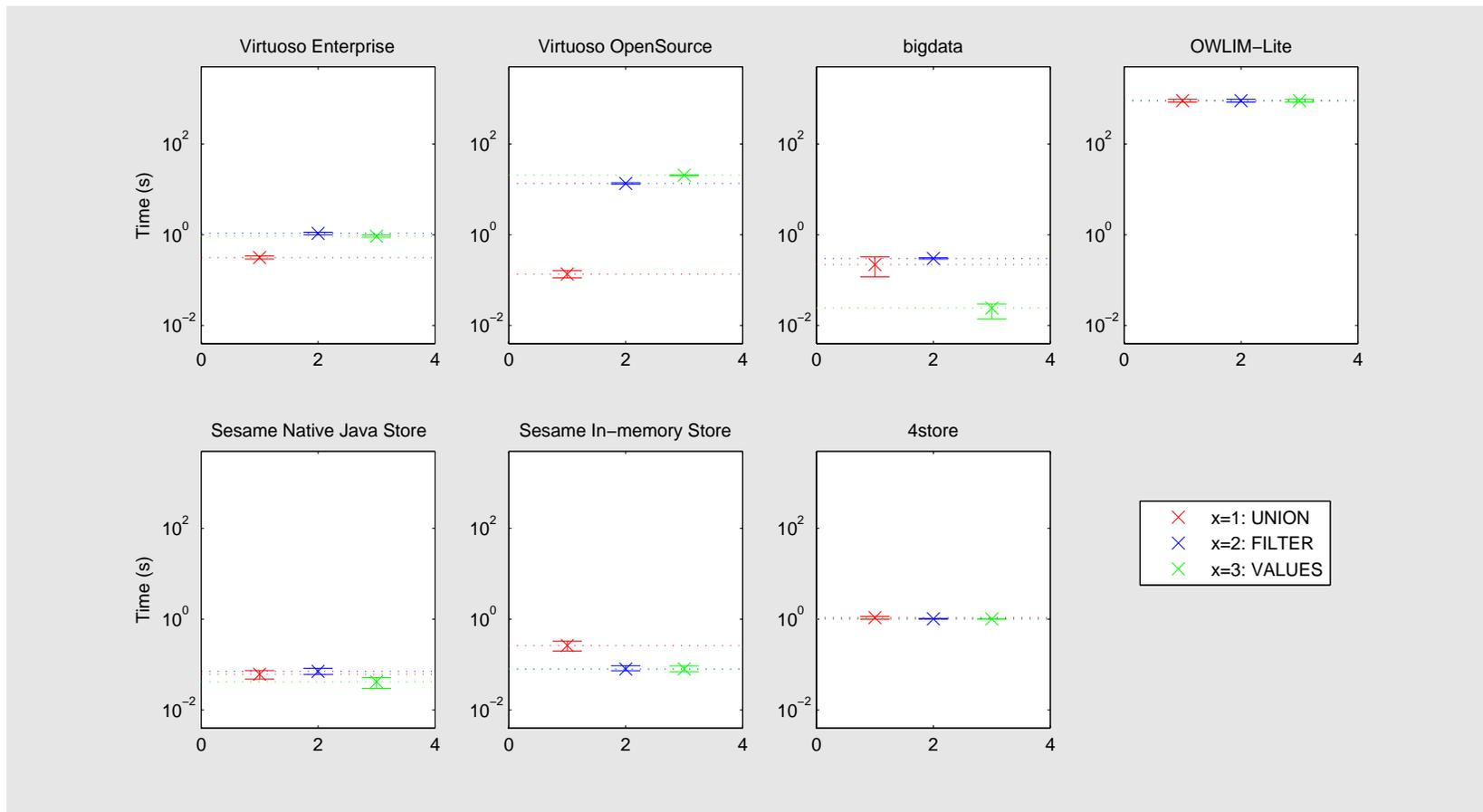}
\caption{Specifying 10 alternative URIs. The average response time for the sample of 500 compounds specified in sets of 10 is marked with $\times$. The error bars indicate the maximum and minimum response times obtained. \label{fig:alternatives_10}}
\end{sidewaysfigure*}

\begin{sidewaysfigure*}
\centering
\includegraphics[width=\textwidth]{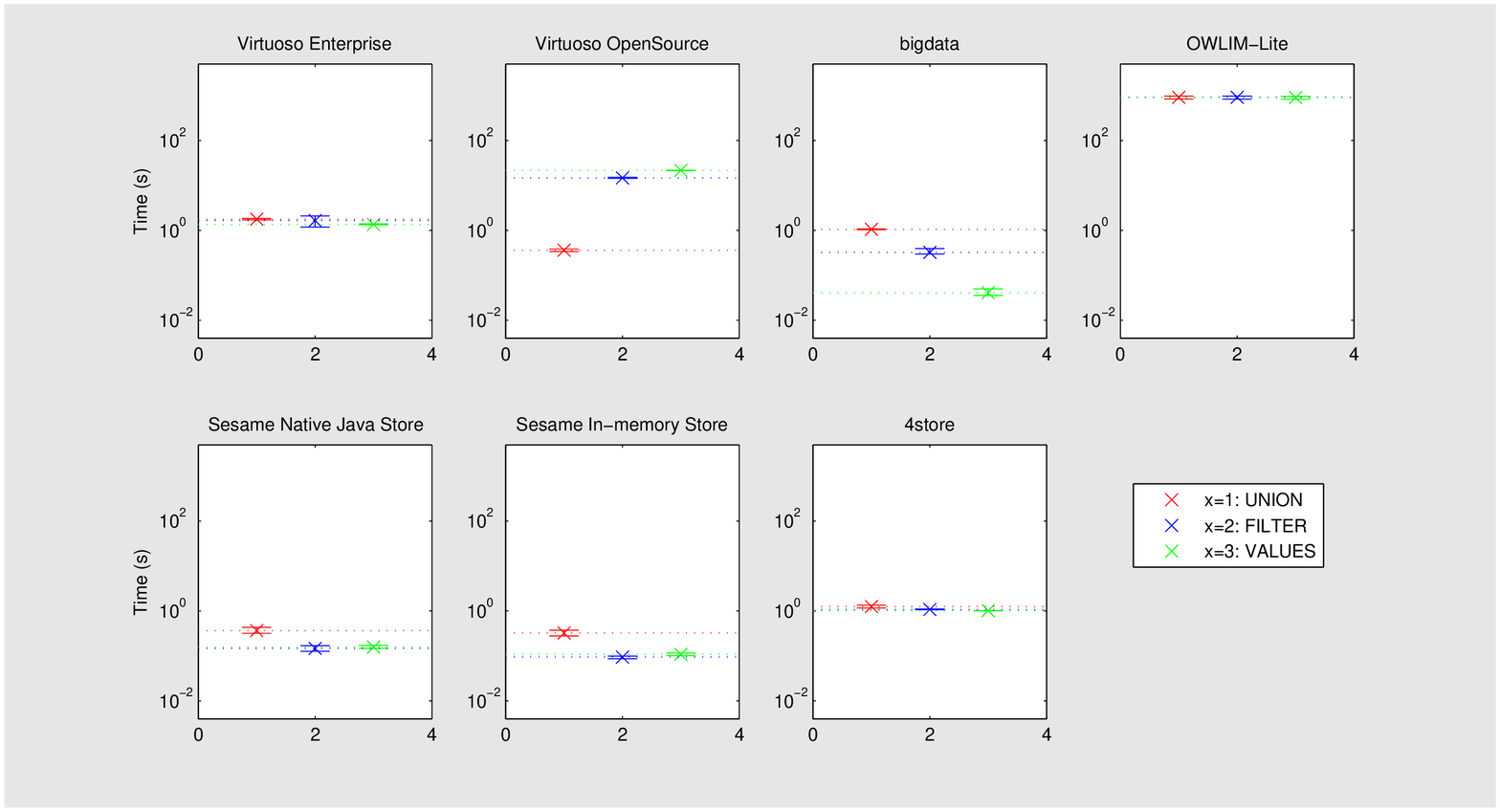}
\caption{Specifying 20 alternative URIs. The average response time for the sample of 500 compounds specified in sets of 10 is marked with $\times$. The error bars indicate the maximum and minimum response times obtained. \label{fig:alternatives_20}}
\end{sidewaysfigure*}

\begin{sidewaysfigure*}
\centering
\includegraphics[width=\textwidth]{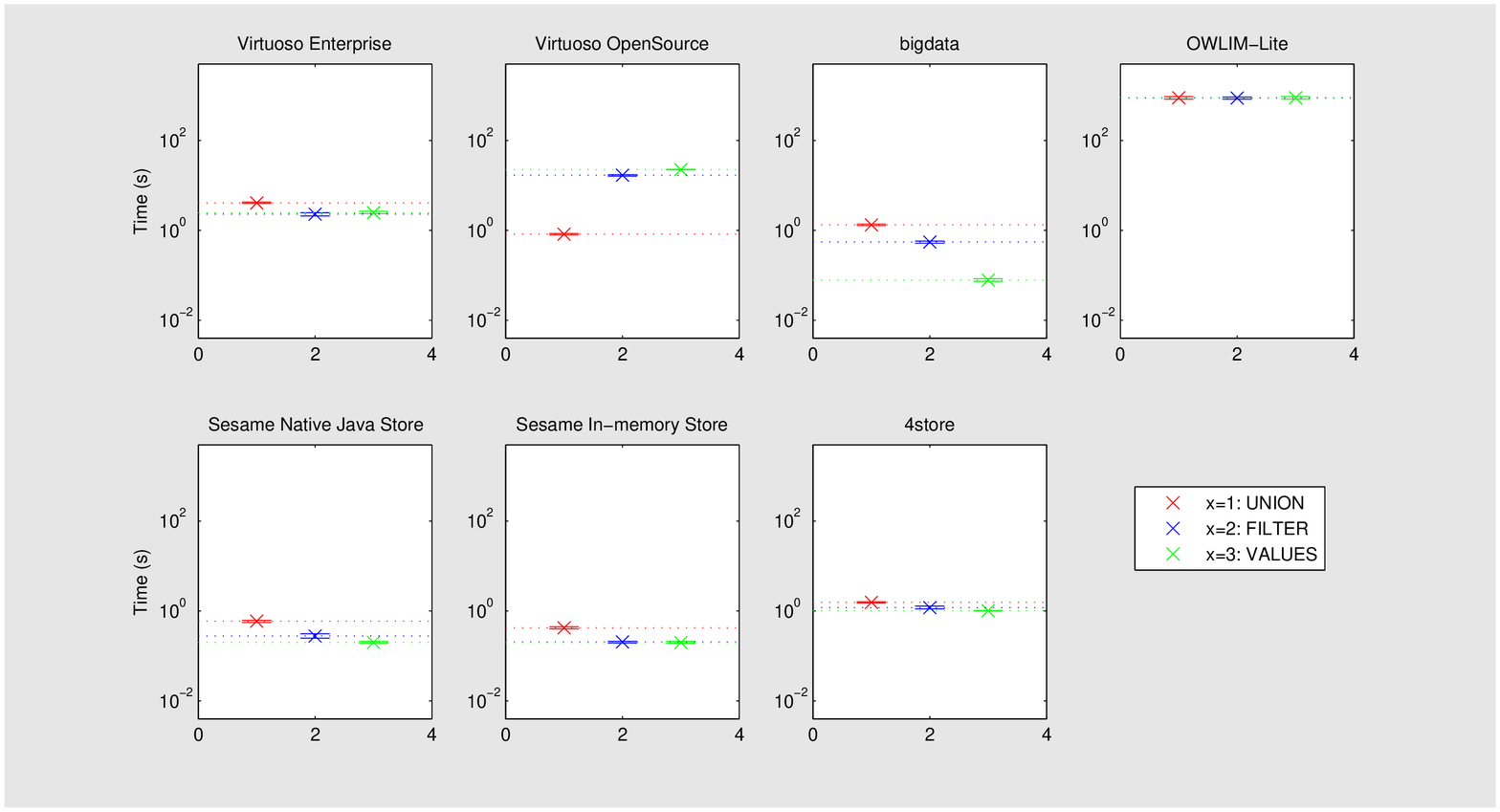}
\caption{Specifying 50 alternative URIs. The average response time for the sample of 500 compounds specified in sets of 10 is marked with $\times$. The error bars indicate the maximum and minimum response times obtained. \label{fig:alternatives_50}}
\end{sidewaysfigure*}

The final set of experiments was carried out to assess the relative performance of the different ways for specifying alternative URIs supported in SPARQL: \texttt{UNION}, \texttt{FILTER}, and \texttt{VALUES}. Three experiments were carried out using subsets of 10, 20, and 50 resources drawn from our initial sample of 500 compounds, while the query used corresponds to the `Compound Pharmacology' method of the OpenPHACTS API. 

The response times across the various RDF stores for 10 alternative URIs are given in Figure \ref{fig:alternatives_10}, Figure \ref{fig:alternatives_20} presents those obtained using 20 alternative URIs and finally Figure \ref{fig:alternatives_50} gives the response times obtained using 50 alternative URIs. In all three figures, the leftmost data point (red $\times$) provides the average response time obtained by specifying the alternative URIs in \texttt{UNION} clauses, the blue $\times$ (middle data point) denotes the mean response time obtained by specifying the alternative URIs using \texttt{FILTER}, and finally the rightmost data point (green $\times$) is the average response time obtained when using \texttt{VALUES}. 

When 10 alternative URIs are specified,  using \texttt{UNION} clauses results in more efficient queries for the two Virtuoso RDF stores,  while there are only small differences in the performance of the three different methods for the Sesame Native Java store and 4store, while \texttt{FILTER} and \texttt{VALUES} provide slightly better response times for the Sesame In--memory store. OWLIM-Lite became corrupted as before, while for bigdata specifying 10 alternative URIs using a \texttt{VALUES} clause is faster than the other two methods by an order of magnitude.

The behaviours of 4store and Virtuoso OpenSource remain stable when the number of alternative URIs is increased to 20, as shown in Figure \ref{fig:alternatives_20}.  However, for all remaining RDF stores (with the exception of OWLIM-Lite), specifying 20 alternative URIs via \texttt{UNION} provides a slower response time than both \texttt{FILTER} and \texttt{VALUES}. In fact, the response times for \texttt{FILTER} and \texttt{VALUES} are very similar to those obtained with only 10 URIs while response times for \texttt{UNION} increase significantly.

Figure \ref{fig:alternatives_50} shows that increasing the number of alternative URIs to 50 does not significantly alter the behaviour of any of the RDF stores we studied with the exception of Virtuoso Enterprise. In this case we observe that the increase in response time as compared to when 20 alternative URIs are used is approximately two time larger for \texttt{UNION} than it is for \texttt{FILTER} or \texttt{VALUES}.

\subsection{Summary of results}

The results from the first set of experiments presented in Section \ref{sec:ex:optimise} provide empirical evidence that formal results published in the literature regarding the use of \texttt{OPTIONAL} graph patterns do carry over to practical applications. 
Specifically, the application of a single algorithm to identify redundant \texttt{OPTIONAL} query patterns can yield dramatic improvements on query performance for all of the RDF stores considered. Moreover, the introduction of named graphs has also been shown to be effective in reducing the execution time for the majority of experiments performed.

Next, the experiments presented in Section \ref{sec:ex:sequence} studied whether replacing connected triple patterns with sequence path expressions can speed up query execution.  The results obtained show that the upper bound on response times can be improved by an on order of magnitude in 4 out of the 5 cases where the experiments were run successfully.

Section \ref{sec:ex:alternatives} presented experiments that compared the performance of three different ways of specifying alternative URIs in SPARQL(\texttt{UNION}, \texttt{FILTER} and \texttt{VALUES}) presenting measurements for 10, 20 and 50 alternative URIs. We found that \texttt{UNION} performs best when 10 URIs are used, but it is outperformed by the other two methods when sets of 20 or 50 URIs were used.

Finally, we note that while some heuristics were found to be ineffective for some query and RDF store combinations, only the application of the \texttt{GRAPH} heuristic has had a negative effect on query execution time, and only in 3 out of the 24 successful experiments.
\section{Practical implications for providing paginated RDF views}\label{sec:practical}

We now describe how these heuristics can be applied in practice for the common use-case of pagination. In many cases, the number of results obtained through the execution of a query is large and can become overwhelming to users if presented all at once. Client side applications are not well placed to deal with this issue, as the processing and pagination of large result sets poses a significant overhead that can severely impact the usability of the application. Thus, a pagination mechanism for SPARQL result sequences is desirable. In this context, a page is considered equivalent to an ordered list of a predefined number of individual results from the same result sequence.

Intuitively, this issue can be dealt with through the use of SPARQLs \texttt{LIMIT} and \texttt{OFFSET} keywords. In practice, applications typically require the ability to change the sort order, and to apply arbitrary filtering rules to control which results from the sequence are displayed in each page, thus posing additional challenges to a server side implementation.

This section illustrates how the heuristics proposed in this paper can be applied to enable the provision of RDF paginated views.

\paragraph{Minimise optional triple patterns and localise sub--patterns}
Assuming the graph patterns that retrieve the required information are known, their conjunction provides an `Initial query'. The heuristics presented in sections \ref{sec:optional} and \ref{sec:named} can then be applied to the initial query to improve its performance.

\paragraph{Eliminate cartesian products}
Subsequently, any results that appear due to cartesian products of variable mappings must be eliminated from sequences intended for pagination in order to ensure that items that appear in a page correspond to individual and independent data points. This is of particular importance to scientific applications since cartesian products can artificially increase the number of results returned by a query, resulting in an overestimation of the number of datapoints recorded in a dataset.
%
%However, the use of aggregates introduces some additional complications. To quote \cite{sparql1.1} ``In aggregate queries and sub-queries, variables that appear in the query pattern, but are not in the \texttt{GROUP BY} clause, can only be projected or used in select expressions if they are aggregated.'' That is, any variables that are not aggregated must appear inside a \texttt{GROUP BY} clause. In turn, any sorting specified via an \texttt{ORDER BY} clause will only be applied inside the generated groups, while the result sequence as a whole will remain in arbitrary order. 
%
%An effective way to sort the entire result sequence is through the use of a sub--query which contains only aggregated variables, and thus does not require grouping, which means the sort order can be specified at the outermost query and will be applied to the entire result sequence.

\paragraph{Specify alternative URIs}
In addition, when each result consists of mappings for a large number of variables it can become difficult to assign meaning to each attribute displayed based solely on the names of variables. Arguably the semantics can be retrieved by considering the graph patterns that have generated the result set. However this approach restricts the result semantics to those of the original schema, and can be inappropriate in a data integration setting. Instead, an RDF projection of the mappings through the use of a \texttt{CONSTRUCT} clause can provide both flexibility on the representation of results and clarity with respect to their semantics.

The provision of RDF paginated view poses further challenges. However, since ordered lists are represented in RDF through a collection of pairs consisting of a value, and a pointer to the next pair through the use of the \texttt{rdf:first} and \texttt{rdf:rest} predicates. As the \texttt{rdf:first} predicate has to be iteratively applied to individual mappings for the same variable, SPARQL RDF projections can not be used to generate such structures. Thus, to achieve pagination of RDF projections a two--step process is required. First, the URIs for items that a page consists of must be identified through a \texttt{SELECT} query so that they can be retrieved in the desired order. These are used to generate the page template through \texttt{rdf:first} and \texttt{rdf:rest}. Subsequently, a \texttt{CONSTRUCT} projection query is issued to obtain the required attributes for each item. To do so, this query must specify alternative URIs, each corresponding to one item retrieved in the previous step. The resulting RDF can then be appended to the page template and forwarded to the client. 

\paragraph{Sorting aggregate result sequences}

The use of aggregates introduces some additional complications. To quote \cite{sparql1.1} ``In aggregate queries and sub-queries, variables that appear in the query pattern, but are not in the \texttt{GROUP BY} clause, can only be projected or used in select expressions if they are aggregated.'' That is, any variables that are not aggregated must appear inside a \texttt{GROUP BY} clause. In turn, any sorting specified via an \texttt{ORDER BY} clause will only be applied inside the generated groups, while the result sequence as a whole will remain in arbitrary order. 

An effective way to sort the entire result sequence is through the use of a sub--query which contains only aggregated variables, and thus does not require grouping. The sort order can then be specified at the outermost query and will be applied to the entire result sequence.

Finally, in order to enable results to be sorted and filtered arbitrarily with respect to the values mapped to each of the variables, the graph patterns that retrieve these values must also appear in the first \texttt{SELECT} query so that the correct URIs are available for use in the \texttt{CONSTRUCT} query.

Here, one can see how these heuristics can be applied in practice. 

\section{Conclusions}\label{sec:conclusions}

This paper presented a set of 5 heuristics that can be used to guide the formulation of performant SPARQL queries. These heuristics were  inspired by formal results found in the literature as well as hands on experience in developing an end-user focused data integration system. These heuristics are proposed as a first step towards helping developers formulate SPARQL queries that are more in-line with the capabilities of state--of--the--art RDF stores. In addition, we hope these heuristics can help RDF store developers to further optimise their stores.

The heuristics were first formally defined in Section \ref{sec:method} and subsequently evaluated in Section \ref{sec:eval}, using openly available real world data and queries used in the OpenPHACTS project. While the results show performance improvements obtained through the application of the heuristics in most cases, it is important to note that there is a large degree of variability. With that in mind, the only instances of a heuristic having a negative impact on query performance concern the introduction of named  graphs to a query with no \texttt{OPTIONAL} clauses which in our experience rarely occurs. Based on these results, we argue that the heuristics presented herein can provide a valuable tool in formulating performant SPARQL queries.

Moreover, the provision of paginated RDF views was considered as a common place application scenario where the application of the heuristics is beneficial. A number of challenges have been identified and solutions based on the heuristics have been proposed for this common use-case.

The large degree of variability observed both across different RDF stores and individual queries provide strong motivation to iteratively test and measure response times for queries considered to drive application requirements. SPARQL, due to its expressiveness , provides a plethora of different ways to express the same constraints,  thus, developers need to be aware of the performance implications of the combination of query formulation and RDF Store. This work provides empirical evidence that can help developers in designing queries for their selected RDF Store. However, this raises questions about the effectives of writing complex {\em generic} queries that work across open SPARQL endpoints available in the Linked Open Data Cloud.  We view the optimisation of queries independent of underlying RDF Store technology as a critical area of research to enable the most effective use of these endpoints. 

Further, future work includes the identification of further heuristics for the formulation of performant SPARQL queries and the study of properties that these queries share. Moreover, we intend to use the OpenPHACTS datasets and queries in the creation of Linked Data benchmarks in the context of the Linked Data Benchmarking Council project.

\subsection*{Acknowledgements}

The research leading to these results has received support from the Innovative
Medicines Initiative Joint Undertaking under grant agreement number 115191,
resources of which are composed of financial contribution from the European
Union's Seventh Framework Programme (FP7/2007- 2013) and EFPIA companies' in
kind contribution.

This work was partially supported by EU project LDBC (FP7-317548).

\bibliographystyle{elsarticle-num-names}

\bibliography{references}

\end{document}